\documentclass[onecolumn,usenatbib]{mnras}

\usepackage{graphicx}	
\usepackage{amsmath}	
\usepackage{amssymb}	

\title[Gamma-Ray Burst with Decaying Turbulence]
{Stochastic Acceleration Model of Gamma-Ray Burst with Decaying Turbulence}

\author[Asano \& Terasawa]{
Katsuaki Asano\thanks{E-mail: asanok@icrr.u-tokyo.ac.jp}
and Toshio Terasawa
\\
Institute for Cosmic Ray Research, The University of Tokyo,
5-1-5 Kashiwanoha, Kashiwa, Chiba 277-8582, Japan
}

\date{Accepted XXX. Received YYY; in original form ZZZ}

\pubyear{2015}

\begin{document}
\label{firstpage}
\pagerange{\pageref{firstpage}--\pageref{lastpage}}
\maketitle

\begin{abstract}
The spectral shape of the prompt emissions of gamma-ray bursts (GRBs)
is typically expressed by the Band function: smooth joining of
two power-law functions for high- and low-energy regions.
To reveal the origin of the Band function, we revisit the stochastic acceleration
model, in which electrons are accelerated via scattering with
turbulent waves in the GRB outflow.
The balance between the acceleration and synchrotron cooling
yields a narrow energy-distribution similar to the Maxwellian
distribution.
The synchrotron spectrum becomes consistent with
the observed hard photon index for the low-energy region.
On the other hand, the narrow electron energy distribution
contradicts the power-law spectrum for the high-energy region.
We consider an evolution of the electron energy distribution
to solve this problem.
The turbulence and magnetic field induced by a certain hydrodynamical instability
gradually decay.
According to this evolution, the typical synchrotron photon energy also decreases with time.
The time-integrated spectrum forms the power-law shape for the high-energy region.
We discuss the required evolutions of the turbulence and
magnetic field to produce a typical Band function.
Although the decay of the turbulence is highly uncertain,
recent numerical simulations for decaying turbulence
seem comparatively positive for the stochastic acceleration model.
Another condition required to reconcile observations is a much shorter duration of
the stochastic acceleration than the dynamical time-scale.
\end{abstract}




\section{Introduction}

Gamma-ray bursts (GRBs) release isotropically equivalent energy
of $E_{\rm iso}=10^{51}$--$10^{54}$ erg as gamma-rays
within a time-scale of $0.1$--$100$ s.
The gamma-rays are considered to be emitted from
relativistic jets with bulk Lorentz factor $\Gamma>100$.
The photon spectra are expressed with the conventional
Band function \citep{ban93} with the parameters of
the spectral peak energy $\varepsilon_{\rm p}$
(typically 100 keV--1 MeV), low-energy photon index $\alpha$
($\sim -1$), and high-energy photon index $\beta$ (roughly between
$-2$ and $-3$).
The short variability ($1$ ms--$1$ s) of GRB pulses
is attributed to intermittent ejections of multiple emission regions
from the central engine.
In the classical internal shock model, this emission regions
are shocked shells, where electrons are accelerated via
the first-order Fermi acceleration (Fermi-I).
However, the observationally typical low-energy index contradicts
this scenario.
Synchrotron emission from electrons injected via the Fermi-I process
yields $\alpha=-1.5$, because the strong synchrotron cooling
leads to an electron energy distribution of $N(E_{\rm e}) \propto E_{\rm e}^{-2}$.

Alternative ideas to resolve the low-energy photon index problem are
photosphere models \citep[and references therein]{mes00,gia06,pee06,asa13}, 
the Klein-Nishina effect on synchrotron self-Compton (SSC)
process \citep[and references therein]{der01,bos09,nak09,wan09,bos14},
sudden damping of magnetic fields \citep{pee06b,zha14},
and inverse Compton models \citep{ghi99,ste04,vur09}.
In this paper, we revisit the second order Fermi acceleration (Fermi-II) model
of \citet{asa09} \citep[see also][]{mur12}.
The Fermi-II process continuously accelerates electrons
during the photon emission period.
In this process, the resultant electron spectral index becomes
harder than $-2$ \citep[e.g.][]{sch84,par95,bec06,sta08,lef11}.
As shown in \citet{asa09}, the acceleration time-scale
required to balance with the synchrotron cooling is much longer than
the shock acceleration time-scale with the Bohm limit;
in other words, the electron mean free path in the Fermi-II models
is much longer than the Larmor radius $r_{\rm L}$.
The scattering efficiency required in the Fermi-II models is not so high
compared to the Fermi-I acceleration.
\citet{asa09} showed that the Fermi-II
model naturally reproduces
the low-energy photon index of the GRB spectrum.
The required turbulence to accelerate electrons
may be induced via the Kelvin--Helmholtz instability
in the shear flow \citep[e.g.][]{zha03,miz09}
or at the boundary between the jet and cocoon \citep{mes01,ram02}.
As radial modes, the Rayleigh--Taylor and Richtmyer--Meshkov instabilities
are possible candidates to induce turbulence \citep{mat13}.
In internal shock simulations with density fluctuation,
the Richtmyer--Meshkov instability is excited \citep{ino11}.

Interestingly, recent studies with the Fermi-II process
also agrees with broadband blazar spectra \citep{asa14,dil14,kak15,asa15}.
The models reasonably reproduce very hard spectra
seen in 1ES 0229+200 and 3C 279, and curved spectra for Mrk 421 and Mrk 501.
Such results encourage us considering the Fermi-II process in GRB jets as well.

While numerical simulations of the Fermi-II were done in \citet{asa09}
and \citet{mur12}, we analytically discuss the stochastic acceleration
and emission in this paper.
Whereas a simple Fermi-II model can easily produce
observationally typical value of $\alpha \sim -1$,
we mainly focus on the high-energy slope of the photon spectrum.
The simplest Fermi-II model implies a Maxwellian-like
narrow energy distribution for electrons,
which contradicts the power-law photon spectrum above $\varepsilon_{\rm p}$.
To resolve this problem,
we consider temporal evolutions of the electron energy distribution,
which may be regulated by the evolutions of the turbulence,
mean magnetic field, and electron injection process.
The resultant evolution of the photon peak energy
would produce a power-law envelope curve in the photon spectrum.
Our purpose in this paper is to find ideal evolutions
of the model parameters to reproduce the GRB photon spectrum.
Although the lack of knowledge with the turbulence property and evolution
in GRB jets prevents the final confirmation of the obtained results,
our study will be an important step to probe the possibility
of the Fermi-II process in GRB.

\section{Stochastic Acceleration and Cooling}
\label{sec:stacc}

\begin{figure}
	\includegraphics[width=\columnwidth]{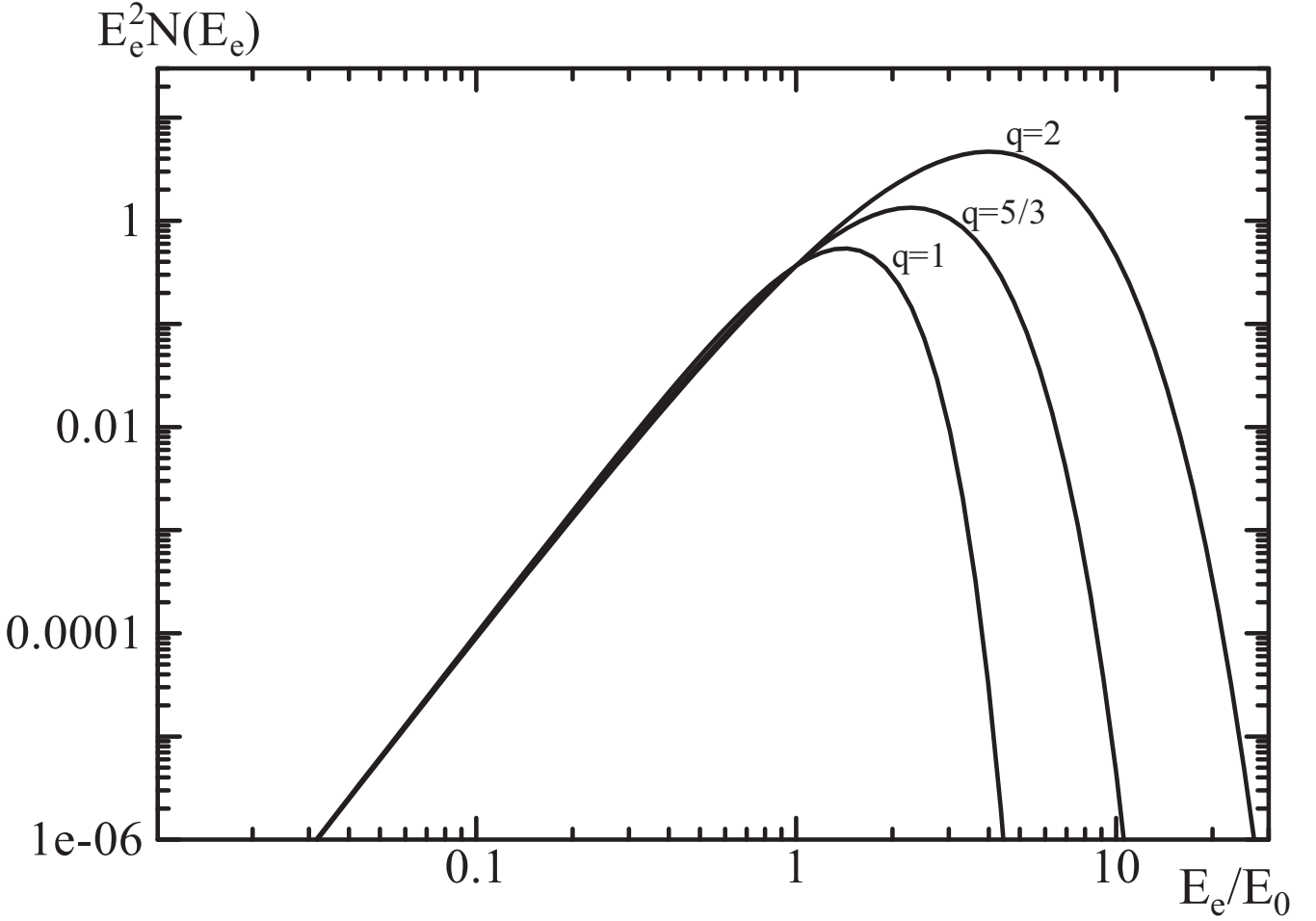}
    \caption{Steady solutions of electron energy distributions $N(E_{\rm e})$
for various $q$.}
    \label{fig:Ne}
\end{figure}

Gamma-rays are emitted from a collimated outflow relativistically
moving towards observers.
Hereafter, we discuss the electron acceleration and photon emission
in the plasma rest frame.
The values in the following equations are defined in the rest frame
except for photon energy; $\varepsilon'$ in the rest frame
and $\varepsilon=\Gamma \varepsilon'$ in the observer frame.
We start from the same Fokker--Planck equation in \citet{asa09} as
\begin{eqnarray}
\frac{\partial N}{\partial t}=\frac{\partial}{\partial E_{\rm e}}
D_{EE} \frac{\partial N}{\partial E_{\rm e}}
-\frac{\partial}{\partial E_{\rm e}}
\left[ \left(2 \frac{D_{EE}}{E_{\rm e}}
-\dot{E}_{\rm cool} \right) N \right],
\label{FP}
\end{eqnarray}
for the energy distribution of ultrarelativistic electrons $N(E_{\rm e})$
in a certain volume.
The energy diffusion coefficient $D_{EE}$
is phenomenologically assumed to have a power-law form
\begin{eqnarray}
D_{EE}=K_0 E_{\rm e}^q.
\end{eqnarray}
For simplicity, we consider only synchrotron radiation
for the cooling processes for electrons.
The energy loss rate for an electron of $E_{\rm e}=\gamma_{\rm e} m_{\rm e} c^2$
is expressed as
\begin{eqnarray}
\dot{E}_{\rm cool}(\gamma_{\rm e})=\frac{4}{3} \sigma_{\rm T} c \gamma_{\rm e}^2
U_B,
\label{cool}
\end{eqnarray}
where $\sigma_{\rm T}$ is the Thomson cross section,
and $U_B \equiv B^2/8 \pi$ is the energy density of the magnetic field.
The cooling time
\begin{eqnarray}
t_{\rm c}(\gamma_{\rm e})=\frac{6 \pi m_{\rm e} c}{\sigma_{\rm T} B^2 \gamma_{\rm e}},
\label{tc}
\end{eqnarray}
for gamma-ray emitting electrons is much shorter than the dynamical time-scale
\citep[see e.g.][]{asa09}.
Therefore, when the synchrotron cooling balances with the stochastic acceleration,
the electron energy distribution can be approximated by the steady solution
for eq. (\ref{FP}) with a pile-up feature \citep{sch84,sch85,aha86,sta08,lef11} as
\begin{eqnarray}
N(E_{\rm e})=\frac{3 N_{\rm tot}}
{\Gamma \left( \frac{6-q}{3-q} \right) E_0}
\left( \frac{E_{\rm e}}{E_0} \right)^2
\exp \left[ -\left( \frac{E_{\rm e}}{E_0} \right)^{3-q} \right],
\label{eq:ele}
\end{eqnarray}
where $N_{\rm tot}$ is the total electron number,
$\Gamma$ is the gamma function, and the cut-off energy is written as
\begin{eqnarray}
E_0=\left( \frac{6 \pi (3-q) K_0 m_{\rm e}^2 c^3}{\sigma_{\rm T} B^2}
\right)^{1/(3-q)}.
\end{eqnarray}
As shown in Figure \ref{fig:Ne},
the spectral peak energy in $E_{\rm e}^2 N(E_{\rm e})$-plot
is sensitive to the index $q$.

\begin{figure}
	\includegraphics[width=\columnwidth]{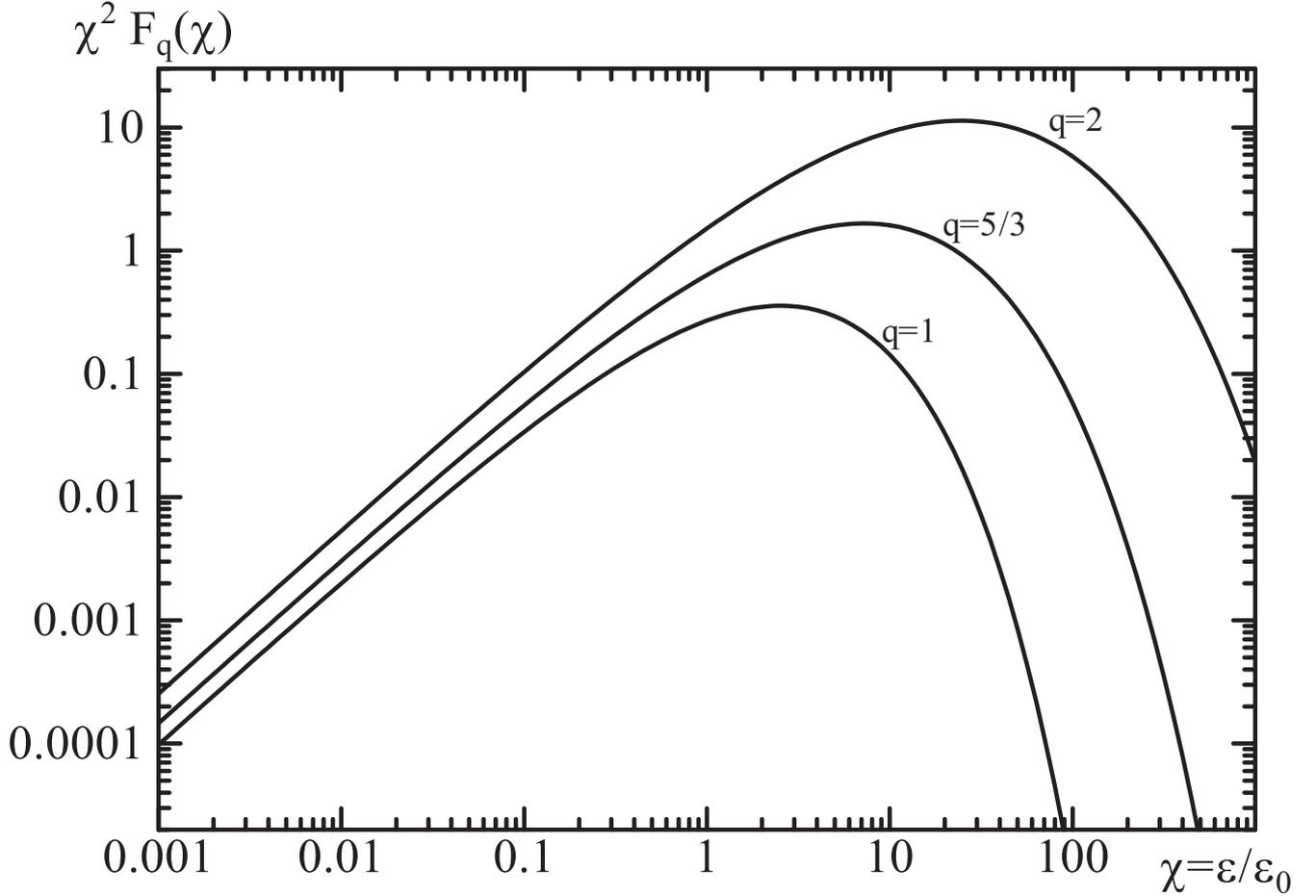}
    \caption{The non-dimensional functions $F_q$
that provide the spectral shape of the photon production rate
$\dot{N}_\gamma$.}
    \label{fig:ndot}
\end{figure}

Adopting the synchrotron function ${\cal F}_{\rm syn}(\chi)
\equiv \chi \int_\chi^\infty d \xi K_{5/3}(\xi)$,
where $K_{5/3}(\xi)$ is the modified Bessel function of the second kind
\citep{ryb79},
the photon production rate can be written as
\begin{eqnarray}
\dot{N}_\gamma(\varepsilon')=\frac{\sqrt{3} e^3 B}{8 \hbar m_{\rm e} c^2 \varepsilon'}
\int dE_{\rm e} N(E_{\rm e})
{\cal F}_{\rm syn} \left( \frac{\varepsilon'}{\varepsilon'_{\rm typ}} \right),
\label{ndot}
\end{eqnarray}
where the typical photon energy for electrons of $E_{\rm e}$ is
\begin{eqnarray}
\varepsilon'_{\rm typ}(E_{\rm e})=\frac{3 \pi \hbar e B}{8 m_{\rm e} c}
\gamma_{\rm e}^2.
\label{eq:typ}
\end{eqnarray}
Substituting eq. (\ref{eq:ele}) into equation (\ref{ndot}),
\begin{eqnarray}
\dot{N}_\gamma(\varepsilon')
&=&\frac{\sqrt{3} e^2 m_{\rm e}^2 c^3 N_{\rm tot}}
{\pi \Gamma \left( \frac{6-q}{3-q} \right) \hbar^2 E_0^3}
\int dE_{\rm e} \exp
\left[ -\left( \frac{E_{\rm e}}{E_0} \right)^{3-q} \right]
\int_{\varepsilon'/\varepsilon'_{\rm typ}}^\infty d \xi
K_{5/3} (\xi).
\end{eqnarray}
Defining $x \equiv E_{\rm e}/E_0$ and $\varepsilon'_0
\equiv \varepsilon'_{\rm typ}(E_0)$,
\begin{eqnarray}
\dot{N}_\gamma(\varepsilon')&=&\frac{\sqrt{3} e^2 m_{\rm e}^2 c^3 N_{\rm tot}}
{\pi \Gamma \left( \frac{6-q}{3-q} \right) \hbar^2 E_0^2}
\int dx \exp \left( -x^{3-q} \right)
\int_{x^{-2} \varepsilon'/\varepsilon'_0}^\infty d \xi
K_{5/3} (\xi) \\
&\equiv&\frac{\sqrt{3} e^2 m_{\rm e}^2 c^3 N_{\rm tot}}
{\pi \Gamma \left( \frac{6-q}{3-q} \right) \hbar^2 E_0^2}
F_q \left(\frac{\varepsilon'}{\varepsilon'_0} \right).
\label{ndot2}
\end{eqnarray}
Then, the spectral shape of the photon production rate
is expressed by the non-dimensional function
$F_q (\varepsilon/\varepsilon_0 )$,
which is obtained by integrating the function over $x$ and $\xi$.
In Figure \ref{fig:ndot}, we show the functional shape of $F_q$
for various $q$.
The horizontal axis is normalized by $\varepsilon_0$.
As shown in Figure \ref{fig:ndot},
the spectral peaks in $\varepsilon^2 \dot{N}_\gamma(\varepsilon)$-plot
are larger than $\varepsilon_0$ \citep{fri89,zir07,lef11}.
Here, we define this peak as $\varepsilon_{\rm p} \equiv X_q \varepsilon_0$,
and summarize values of $X_q$ in Table~\ref{tab}.

\begin{table}
	\centering
\caption{$X_q \equiv \varepsilon_{\rm p}/\varepsilon_0$,
the non-dimensional function $I_q(0,0)$ is defined
by eq. (\ref{eq:I}).}
\label{tab}
\begin{tabular}{lcccc}
\hline
$q$ & $1$ & $3/2$ & $5/3$ & $2$   \\ \hline
$X_q$ & 2.5 & 5.0 & 7.2 & 25 \\
$I_q(0,0)$ & 1.1 & 3.0 & 5.3 & 35 \\ \hline
\end{tabular}
\end{table}

We can choose fiducial parameters to agree with $\varepsilon_{\rm p} \sim 0.5$ MeV as
\begin{eqnarray}
\varepsilon_{\rm p}=
1.0 \left( \frac{\Gamma}{500} \right) \left( \frac{K_0}{10^{2}~\mbox{s}^{-1}} \right)^2
\left( \frac{B}{10^4~\mbox{G}} \right)^{-3}~\mbox{MeV},&&~\mbox{for}~q=2,
\label{eq:epfirst} \\
\varepsilon_{\rm p}=
0.72 \left( \frac{\Gamma}{500} \right) \left( \frac{K_0}{10^{2}~\mbox{GeV}^{1/3}
~\mbox{s}^{-1}} \right)^{3/2}
\left( \frac{B}{10^4~\mbox{G}} \right)^{-2}~\mbox{MeV},&&~\mbox{for}~q=5/3,\\
\varepsilon_{\rm p}=
0.65 \left( \frac{\Gamma}{500} \right) \left( \frac{K_0}{10^{2}~\mbox{GeV}^{1/2}
~\mbox{s}^{-1}} \right)^{4/3}
\left( \frac{B}{10^4~\mbox{G}} \right)^{-5/3}~\mbox{MeV},&&~\mbox{for}~q=3/2,\\
\varepsilon_{\rm p}=
0.52 \left( \frac{\Gamma}{500} \right) \left( \frac{K_0}{10^{2}~\mbox{GeV}
~\mbox{s}^{-1}} \right)
\left( \frac{B}{10^4~\mbox{G}} \right)^{-1}~\mbox{MeV},&&~\mbox{for}~q=1.
\label{eq:eplast}
\end{eqnarray}
Note that the typical electron energy to emit 0.1--1 MeV photons
is $\sim$ GeV in the source rest frame
for $\Gamma=500$ and $B=10^4$ G.
The acceleration time-scale $t_{\rm acc} \propto E_{\rm e}^{2-q}$
implied above ($\sim 10^{-2}$ s for GeV electrons)
is much longer than the gyration period $2 \pi E_{\rm e}/(eBc)
\sim 7 \times 10^{-8} (E_{\rm e}/\mbox{GeV}) (B/10^4~\mbox{G})^{-1}$ s.
The Fermi-II model implies that the energy gain per one scattering
is $\Delta E_{\rm e}/E_{\rm e} \sim \beta_{\rm eff}^2$,
where $\beta_{\rm eff}$ is the effective turbulence velocity normalized by $c$.
The energy fraction of the turbulence to the bulk jet energy
is $\sim \beta_{\rm eff}^2$.
Though the turbulence velocity may be slower than the sound velocity
in the relativistic plasma ($\beta_{\rm eff}^2<1/3$),
the high radiative efficiency \citep[e.g.][]{zha07}
may require $\beta_{\rm eff}^2 \sim 0.1$.
This situation is similar to the standard shock acceleration models,
in which a mildly relativistic shock is presumed.
In both the cases, the energy source of the non-thermal electrons
is the energy of mildly relativistic protons.
Here, we assume that the turbulence has a characteristic scale,
namely typical eddy size $l_{\rm edd}$.
If the turbulence spectrum ranges in significantly short wavelengths
comparable to the Larmor radius $r_{\rm L} \sim 300 (E_{\rm e}/10^9~\mbox{eV})
(B/10^4~\mbox{G})^{-1}$,
the mean free path is written as
$l_{\rm mfp}\sim r_{\rm L} B^2/\delta B_{\rm eff}^2$ \citep{bla87}.
Inside the eddies $\beta_{\rm eff}^2 \ll 0.1$,
while only electrons moving across eddies are scattered
via nonresonant mirror interaction
(transit time damping) arose from compressible (acoustic) modes
\citep{ptu88,cho06,yan08,lyn14} with $\beta_{\rm eff}^2 \sim 0.1$.
When the diffusion time $l_{\rm edd}^2/(c l_{\rm mfp})$
is much longer than the eddy time-scale $t_{\rm edd}=l_{\rm edd}/(c \beta_{\rm eff})$,
the time-scale $t_{\rm acc} \beta_{\rm eff}^2$ may correspond to the eddy time-scale
so that $l_{\rm edd} \sim c t_{\rm acc} \beta_{\rm eff}^3 \sim 10^7
(t_{\rm acc}/10^{-2}~\mbox{s}) (\beta_{\rm eff}^2/0.1)^{3/2}~\mbox{cm}$.
Finally we obtain $D_{EE} \sim \beta_{\rm eff}^2 E_{\rm e}^2
/t_{\rm edd} \propto \beta_{\rm eff}^3/l_{\rm edd}$.

However, there may be a transition scale at the wavelength where
the kinetic energy is comparable to the magnetic energy.
At the shorter wavelength than the transition scale,
the power-spectrum of the turbulence is steeply damped.
Then, the mean free path of electrons may be elongated
as long as the eddy scale.
In this case, we obtain $l_{\rm edd} \sim c t_{\rm acc} \beta_{\rm eff}^2 \sim
3 \times 10^7
(t_{\rm acc}/10^{-2}~\mbox{s}) (\beta_{\rm eff}^2/0.1)~\mbox{cm}$
and $D_{EE} \propto \beta_{\rm eff}^2/l_{\rm edd}$.

The above qualitative discussion is still insufficient to conclude
the acceleration property in the turbulence.
In the following discussion we maintain the phenomenological parametrization
for the energy diffusion coefficient with arbitrary $q$ and $K_0$.

\section{Decaying Turbulence}
\label{sec:turb}

If the parameters in equation (\ref{ndot2}), such as $\varepsilon_0$ etc.,
remain constant during the emission period,
the photon spectrum for observers is just a blue-shifted spectrum
of $\dot{N}_\gamma(\varepsilon')$.
As shown in Figure \ref{fig:ndot}, the spectral shape with
a sharp cut-off may contradict the observed Band spectra
especially for smaller $q$.
However, the efficiency of the stochastic acceleration
is expected to decay with time.
Then, the typical photon energy $\varepsilon_0$ would
shift to lower energy.
Such an evolution of the photon emission spectrum
can produce a high-energy power-law part
in the time-integrated spectrum.

Since the time-integrated photon spectrum is written as
\begin{eqnarray}
N_\gamma(\varepsilon')&=&\int dt
\dot{N}_\gamma(\varepsilon') \\
&=&\frac{\sqrt{3} e^2 m_{\rm e}^2 c^3}
{\pi \Gamma \left( \frac{6-q}{3-q} \right) \hbar^2}
\int dt \frac{N_{\rm tot}}{E_0^2}
F_q \left(\frac{\varepsilon'}{\varepsilon'_0} \right),
\end{eqnarray}
the temporal evolutions of the two quantities,
$\varepsilon_0$ and $N_{\rm tot}/E_0^2$, determine the final
photon spectral shape.
Let us assume that all the parameters evolve
following power-laws of $(t+t_{\rm tbl})$
and the turbulence suddenly disappears at $t=t_{\rm f}$
for simplicity.
The duration $t_{\rm f}$ is defined as the energy transfer time-scale
from the turbulence to electrons.
The parameter $t_{\rm tbl}$, which may correspond to the
production time-scale of the turbulence, is introduced to avoid
divergence at $t=0$.
Hereafter,  the combinations of the parameters
are assumed to behave as $N_{\rm tot}/E_0^2 \propto (t+t_{\rm tbl})^a$
and $\varepsilon'_0 \propto B E_0^2 \propto (t+t_{\rm tbl})^{-b}$
with indices $a$ and $b$.
Here we explore ideal pairs of the phenomenological indices $a$ and $b$
to reproduce typical GRB spectra.

\begin{figure}
	\includegraphics[width=\columnwidth]{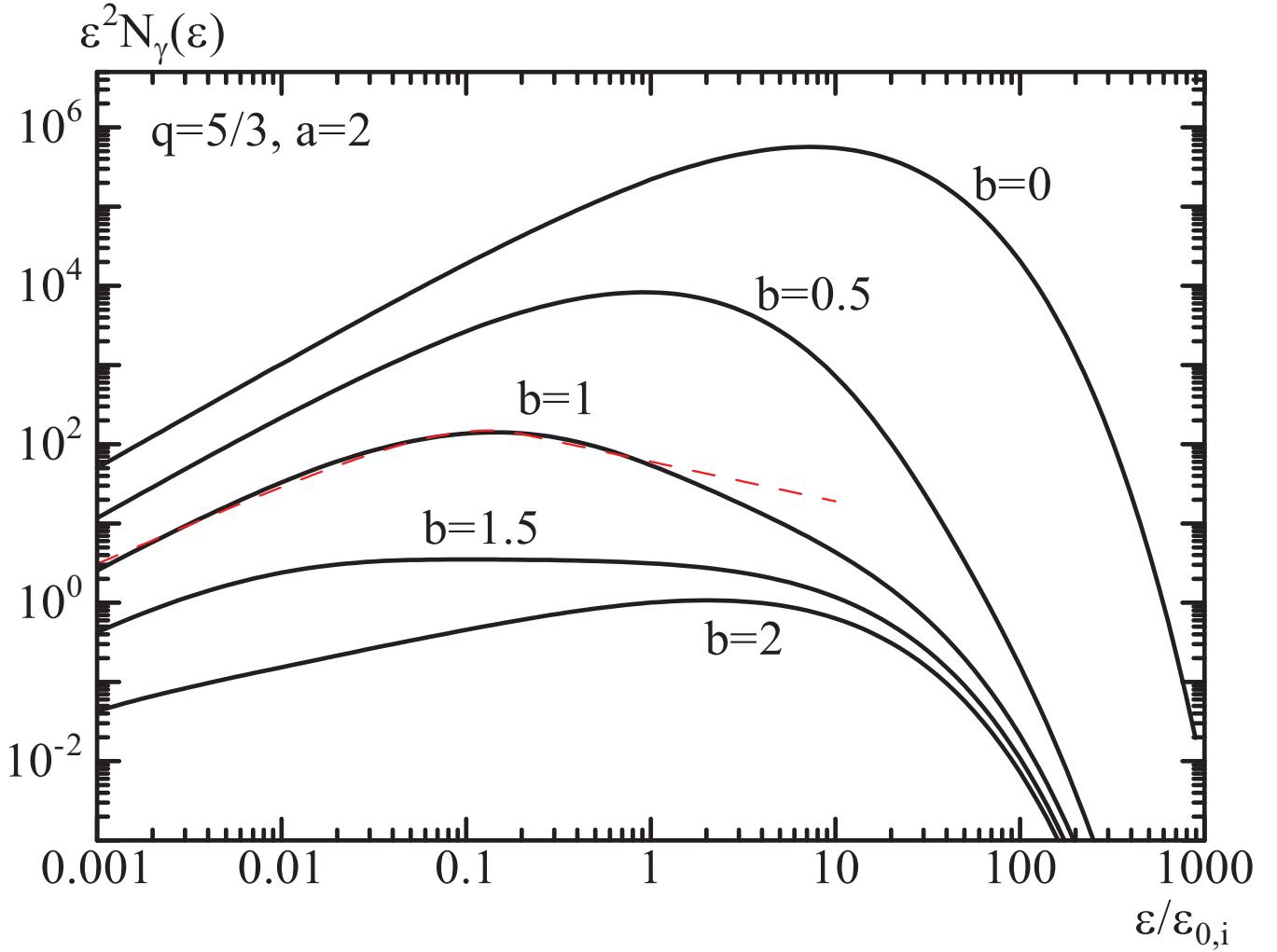}
    \caption{Photon spectrum
with $q=5/3$, $a=2$, and $t_{\rm f}=100 t_{\rm tbl}$ for various $b$.
The dashed line is the Band function with $\alpha=-1$,
$\beta=-2.5$, and $\varepsilon_{\rm p}=0.13 \varepsilon_{\rm 0,i}$.}
    \label{fig:spec}
\end{figure}

As will be discussed below, the index $a$ may be $\lesssim 2$.
In Figure \ref{fig:spec}, we plot the photon spectrum $N_\gamma(\varepsilon)$
with $q=5/3$, $a=2$, and $t_{\rm f}=100 t_{\rm tbl}$ for various $b$.
The photon energy is normalized by $\varepsilon_{\rm 0,i}$, which
is $\varepsilon_0$ at $t=0$.
As $b$ becomes large, the faster decay of $\varepsilon_0$ makes
a lower spectral peak energy.
Since $\int dt \dot{N}_\gamma (\varepsilon'_0) \propto (t+t_{\rm tbl})^{1+a}$,
the photon index parameter in the Band spectrum
would be approximated as
\begin{eqnarray}
\beta=-\frac{1+a}{b}.
\end{eqnarray}
Therefore, as shown in Figure \ref{fig:spec},
the parameter values of $b$ between 1 and 1.5
are favorable for $a=2$ to adjust the index between $-2$ and $-3$.
This result is common irrespective of the value of $q$;
the spectral shapes for other $q$ are similar to each other.
Especially, the case of $b=1$ in Figure \ref{fig:spec}
is well fitted with the typical Band parameters
($\alpha=-1$ and $\beta=-2.5$) below
$10 \varepsilon_{\rm p}$.
The obtained Band parameter $\varepsilon_{\rm p}$ is slightly larger than
the final value of $X_q \varepsilon_0=7.2 (t+t_{\rm tbl})^{-1} \varepsilon_{\rm 0,i}
\simeq 0.07 \varepsilon_{\rm 0,i}$,
but the difference may be hard to be seen as shown in the figure.
The spectral curvature around $\varepsilon_{\rm p}$
leads to the difference
between the Band parameter $\beta=-2.5$ below $10 \varepsilon_{\rm p}$
and the asymptotic value $-(1+a)/b=-3$.
Note that, however,
in most cases the photon statistics above $10 \varepsilon_{\rm p}$ (typically a few MeV)
is insufficient to detect the spectral break or curvature.

Here we attempt to associate the phenomenological parameters $a$ and $b$
with the evolutions of the turbulence and electron injection.
Initially the diffusion coefficient and magnetic field
are larger than the values assumed in eqs. (\ref{eq:epfirst})
--(\ref{eq:eplast}).
Then, the acceleration time-scale is gradually elongated,
and settles down to the value suited for $\varepsilon_{\rm p} \sim 0.5$ MeV
at $t=t_{\rm f}$.
Let us write $K_0 \propto t^{-s}$ and $B \propto t^{-w}$,
$E_0 \propto t^{(2w-s)/(3-q)}$.
Then, $\varepsilon'_0 \propto B E_0^2 \propto t^{-w+2(2w-s)/(3-q)}$.
The total number of electrons may increase
following the injection rate
\footnote{Note that the electron number was assumed to be constant in the simulations
of \citet{asa09}}.
If we express this as $N_{\rm tot} \propto t^n$,
$t \dot{N}_\gamma (\varepsilon'_0) \propto t N_{\rm tot} E_0^{-2}
\propto t^{1+n-2(2w-s)/(3-q)}$.
As a result, the parameters $a$ and $b$ are written
with the indices defined above ($s$, $w$, and $n$) as
\begin{eqnarray}
a=n-2 \frac{2w-s}{3-q},~b=w-2 \frac{2w-s}{3-q}.
\end{eqnarray}

Although the parameters $s$, $w$, and $n$ are highly uncertain,
some discussion of the interaction between the turbulence
and electrons may provide insight into the required condition, $(1+a)/b = 2$--$3$,
in our Fermi-II model.
The average energy gain of electrons per one scattering
by waves is proportional to $\beta_{\rm eff}^2 \sim
\max{(\beta_{\rm W}^2, \beta_{\rm A}^2)}$,
where $\beta_{\rm W}$ and $\beta_{\rm A}$ are the turbulence velocity 
and Alfv\'en velocity normalized by $c$, respectively.
In our case, the required electron energy density is
higher than the magnetic field density so that
$\beta_{\rm W}^2>\beta_{\rm A}^2$ is likely.
In the simulations of \citet{ino11},
when a relativistic shock propagates inhomogeneous medium,
the Richtmyer--Meshkov instability is excited
and the turbulence starts to decay after a few eddy-turnover times.
If we adopt those results,
$\beta_{\rm W}^2 \propto U_{\rm K} \propto t^{-1.3}$
and $\delta B^2 \propto U_B \propto t^{-0.7}$.
As we discussed in the last part of \S \ref{sec:stacc},
the diffusion coefficient may behave as
$D_{EE} \propto \beta_{\rm W}^3/l_{\rm edd}$ (short-mfp model) or
$\beta_{\rm W}^2/l_{\rm edd}$ (long-mfp model).
The long-mfp model with a constant $l_{\rm edd}$
and the simulation result of \citet{ino11}
lead to $q=2$, $s=1.3$.
Since the average field $B \propto \delta B$
in this highly turbulent plasma, we obtain $w=0.35$.
The above values imply $b=1.55$ and $a=n+1.2$.
Therefore, the constant injection rate $n=1$ seems favorable,
while the short-mfp model requires an ad hoc value of $n \sim 3$.

Other recent MHD simulations \citep{zra14,bra15}
show inverse transfers from small to large turbulence-scales
in spite of the absence of magnetic helicity.
In such cases, the evolutions of the turbulence scale
and energy densities depend on their initial conditions
\citep{ole97}.
In the relativistic simulation in \citet{zra14},
the longest wavelength evolves as $k_{\rm M} \propto t^{-2/5}$
and $U_B \propto t^{-14/15}$, while
the non-relativistic simulation of \citet{bra15}
shows $k_{\rm M} \propto t^{-1/2}$
and $U_B \propto t^{-1}$.
The indices of the power-spectral density for $k>k_{\rm M}$ are
2 in both the two simulations.
The simplest analytical discussion in \citet{son99},
considering the evolution of patchy turbulence,
concludes $k_{\rm M} \propto t^{-2/5}$
and $U_{\rm K} \propto U_B \propto t^{-6/5}$,
which are not far from the simulation results.
From the derived indices
$w=3/5$ and $s=8/5$ for the long-mfp model,
finally we obtain the phenomenological indices
$a=n+4/5$ and $b=7/5$.
If the electron injection rate is constant ($n=1$),
the photon index becomes an ideal value $\beta=-(1+a)/b=-2$.
This primitive model of the turbulence evolution
also predicts a reasonable range of the photon index.
The electron acceleration affects the decay of turbulence,
which modifies the estimate of the photon index.
More quantitative discussion based on numerical simulations
is necessary to determine the value $\beta$.
On the other hand, the short-mfp model requires $n=3$ again.

Although the above discussion does not always validate the Fermi-II model,
we do not find any discouraging evidence so far.
The next step is a study to verify the fundamental acceleration process
in such turbulence, but this is beyond our scope in this paper.

We have neglected the effect of inverse Compton scattering.
However, most of GRBs do not show the extra component that can be
interpreted as the inverse Compton component \citep{ack12}.
Even when the magnetic field is weak compared to the electron
energy density, the lack of the inverse Compton component can be explained
by the Klein-Nishina effect \citep[e.g.][]{asa11}.
The numerical simulations of \citet{asa09} also shows
that the inverse Compton emission does not greatly affect the synchrotron spectrum.

\section{Secondary Electron--Positron Pairs}
\label{sec:eiso}

The continuous electron acceleration increases
the gamma-ray energy emitted from one electron so that
the required electron number is less than the standard shock
acceleration models, in which all electrons are assumed to be accelerated.
In our model there should be a condition for electrons to enter
the acceleration process. The injection mechanism is uncertain
similarly to that in the shock acceleration.
The electron injection mechanism may be sub-shocks \citep{nar09}
or magnetic reconnection
in the turbulence.
At least, the electron should be relativistic to be in the acceleration
process we have assumed.
Hereafter, we assume the threshold energy of electrons/positrons,
$\gamma_{\rm th} m_{\rm e} c^2$, to be accelerated by turbulence.

We have assumed a gradual increase of the electron number
by a stable electron injection.
However, high energy photons produce secondary electron--positron pairs
via $\gamma \gamma$-absorption.
Those pairs can be also accelerated by turbulence and emit
high-energy photons.
If such pairs dominate the lepton number,
the non-linear growth of the electron/positron number occurs,
which results in a sudden increase of pairs.
Those non-linear evolution of the electron/positron number is not favorable to
produce a Band-like spectrum.
In this section, we estimate the contribution of the secondary pairs
under the situation we have assumed.

First, we obtain the required number of electrons in the Fermi-II model.
In our model, following the power-law evolution of $(1+t/t_{\rm tbl})^{-b}$,
the typical photon energy decays from $\varepsilon_{\rm 0,i}$
to $\varepsilon_{\rm 0,f}$.
Denoting the values of $N_{\rm tot}$ and $E_0$ at $t=t_{\rm f}$
as $N_{\rm tot,f}$ and $E_{\rm 0,f}$, respectively,
the total photon energy in the comoving frame is written as
\begin{eqnarray}
E'_{\rm ph}=\int d \varepsilon' \varepsilon'
N_\gamma(\varepsilon')
=\frac{\sqrt{3} e^2 m_{\rm e}^2 c^3}
{\pi \Gamma \left( \frac{6-q}{3-q} \right) \hbar^2}
\frac{N_{\rm tot,f}}{E_{\rm 0,f}^2}
\left( 1+\frac{t_{\rm f}}{t_{\rm tbl}} \right)^{-a-1} t_{\rm f}
\varepsilon_{\rm 0,i}^{\prime 2} \nonumber \\
\times \int d y
\int_0^{t_{\rm f}/t_{\rm tbl}} dx y (1+x)^a
F_q \left(\frac{y}{(1+x)^{-b}} \right).
\end{eqnarray}
The peak energy for observers is written with the final value of $\varepsilon_0$ as
$\varepsilon_{\rm p} \simeq \Gamma X_q \varepsilon'_{\rm 0,f}$.
Then, we obtain
\begin{eqnarray}
\Gamma \varepsilon'_{\rm 0,i}
=\frac{\varepsilon_{\rm p}}{X_q} \left( 1+\frac{t_{\rm f}}{t_{\rm tbl}} \right)^{b}.
\end{eqnarray}
Substituting $E_{\rm 0,f}/(m_{\rm e} c^2)$ for $\gamma_{\rm e}$ in eq. (\ref{eq:typ}),
we obtain
\begin{eqnarray}
\frac{\varepsilon'_{\rm 0,i}}{E_{\rm 0,f}^2}=\left( 1+\frac{t_{\rm f}}{t_{\rm tbl}} \right)^{b}
\frac{3 \pi \hbar e B_{\rm f}}{8 m_{\rm e} c}
\left( \frac{1}{m_{\rm e} c^2} \right)^2,
\end{eqnarray}
as well, where $B_{\rm f}$ is the final value of $B$.
Adopting those equations, the isotropically-equivalent total photon energy
in one pulse is expressed as
\begin{eqnarray}
E_{\rm ph}=\Gamma \int d \varepsilon' \varepsilon'
N_\gamma(\varepsilon')&=&\frac{3 \sqrt{3} e^3}
{8 \Gamma \left( \frac{6-q}{3-q} \right) \hbar m_{\rm e} c^2} N_{\rm tot,f} B_{\rm f}
t_f \frac{\varepsilon_{\rm p}}{X_q} I_q(a,b),
\end{eqnarray}
where the non-dimensional function
\begin{eqnarray}
I_q(a,b) &\equiv& \left( 1+\frac{t_{\rm f}}{t_{\rm tbl}} \right)^{2b-a-1}
\int d y
\int_0^{t_{\rm f}/t_{\rm tbl}} dx y (1+x)^a
F_q \left(\frac{y}{(1+x)^{-b}} \right) \label{eq:I} \\
&\simeq& \frac{1}{1+a-2b}
\int dy y F_q \left(y \right)=\frac{1}{1+a-2b} I_q(0,0),
\end{eqnarray}
for $1+a-2b>0$, and the final approximation
has been obtained with $t_{\rm f}/t_{\rm tbl} \gg 1$.
The values of $I_q(0,0)$ for various $q$ are tabulated in Table \ref{tab}.

While the observables are $E_{\rm ph}$ and $\varepsilon_{\rm p}$,
the model parameters to determine the total photon energy
are $N_{\rm tot,f}$, $B_{\rm f}$, and $t_{\rm f}$.
The time-scale $t_{\rm f}$ can be considered as the dissipation time-scale
of the turbulence into electrons.
Alternatively, as will be explained below,
the injection of secondary electron--positron pairs may
control the duration $t_{\rm f}$.
The time-scale $t_{\rm f}$ 
can be shorter than the dynamical time-scale
$t_{\rm dyn} \equiv R/(c \Gamma)$ as supposed in \citet{asa09}.
Note $t_{\rm dyn}=\Gamma t_{\rm v}=50 (\Gamma/500) (t_{\rm v}/0.1~\mbox{s})$ s,
where $t_{\rm v} \equiv R/(c \Gamma^2)$ is the variability time-scale for observers.
But $t_{\rm f}$ may be longer than the cooling time-scale
\begin{eqnarray}
t_{\rm c}(\varepsilon_{\rm p}) \simeq \frac{6 \pi}
{\sigma_{\rm T} B^{3/2}}
\sqrt{\frac{\hbar e \Gamma m_{\rm e} c}{\varepsilon_{\rm p}}}
=2.6 \left( \frac{\Gamma}{500} \right)^{1/2}
\left( \frac{B_{\rm f}}{10^4~\mbox{G}} \right)^{-3/2}
\left( \frac{\varepsilon_{\rm p}}{0.5~\mbox{MeV}} \right)^{-1/2}~\mbox{ms}.
\end{eqnarray}
The total photon energy is written as
\begin{eqnarray}
E_{\rm ph} \sim 1.9 \times 10^{51} \left( \frac{N_{\rm tot,f}}{10^{50}} \right)
\left( \frac{t_{\rm f}}{0.1~\mbox{s}} \right)
\left( \frac{B_{\rm f}}{10^4~\mbox{G}} \right)
\left( \frac{\varepsilon_{\rm p}}{0.5~\mbox{MeV}} \right)~\mbox{erg},
\label{eq:totphe}
\end{eqnarray}
for $q=5/3$ with $I_q(0,0)$.
The required electron number $N_{\rm tot,f}$ to achieve
$\sim 10^{51}$ erg is much less than the number in the
classical shock acceleration model by a factor of $\sim t_{\rm c}/t_{\rm f}$.
The assumed time-scale above implies a significantly
longer scale than the eddy scale estimated in \S \ref{sec:stacc}
as $c t_{\rm f} \sim 100 l_{\rm edd}$.

Let us estimate the number of the secondary pairs.
We approximate the photon-density spectrum for $\varepsilon'>\varepsilon'_{\rm p}$
by a power-law
\begin{eqnarray}
n_\gamma(\varepsilon') \simeq \frac{n_0}{\varepsilon'_{\rm p}}
\left( \frac{\varepsilon'}{\varepsilon'_{\rm p}} \right)^\beta.
\label{eq:phd}
\end{eqnarray}
Although the result will be independent of the width of the emission region,
here we assume the width as $c t_{\rm f}/3$.
Then, the normalization in eq. (\ref{eq:phd}) becomes
\begin{eqnarray}
n_0 \simeq 
\frac{-3(\beta+2) E_{\rm ph}}{4 \pi R^2 c t_{\rm f} \varepsilon_{\rm p}}.
\end{eqnarray}
In this case, the optical depth for $\gamma \gamma$-absorption
is written as
\begin{eqnarray}
\tau_{\gamma \gamma}(\varepsilon') \simeq 0.1 \sigma_{\rm T} n_0
\frac{c t_{\rm f}}{3}
\left( \frac{\varepsilon' \varepsilon'_{\rm p}}{m_{\rm e}^2 c^4} \right)^{-\beta-1},
\end{eqnarray}
\citep[][and supporting material in Abdo et al. 2009]{asa03}.
This results in
\begin{eqnarray}
\tau_{\gamma \gamma}(\gamma m_{\rm e} c^2) \simeq 0.1 \sigma_{\rm T}
\frac{-(\beta+2) E_{\rm ph}}{4 \pi R^2 \varepsilon_{\rm p}}
\Gamma^{1+\beta}
\left( \frac{\gamma \varepsilon_{\rm p}}{m_{\rm e} c^2} \right)^{-\beta-1}.
\end{eqnarray}
When $\tau_{\gamma \gamma}(\gamma_{\rm th} m_{\rm e} c^2) \ll 1$,
the number of the secondary pairs above $\gamma_{\rm th} m_{\rm e} c^2$
is independent of $\gamma_{\rm th}$ as
\begin{eqnarray}
N_\pm (\gamma>\gamma_{\rm th})
&\sim& \tau_{\gamma \gamma}(\gamma_{\rm th} m_{\rm e} c^2)
4 \pi R^2 \frac{c t_{\rm f}}{3} \gamma_{\rm th} m_{\rm e} c^2
n'_\gamma(\gamma_{\rm th} m_{\rm e} c^2) \\
&=&0.1
\frac{(\beta+2)^2 \sigma_{\rm T}}{4 \pi c^2 t_{\rm v}^2}
\frac{m_{\rm e} c^2 E_{\rm ph}^2}{\varepsilon_{\rm p}^3}
\Gamma^{2 \beta-2}
\left( \frac{\varepsilon_{\rm p}}{m_{\rm e} c^2} \right)^{-2\beta-1}.
\end{eqnarray}
For $\beta=-2.5$, the number becomes
\begin{eqnarray}
N_\pm (\gamma>\gamma_{\rm th})
&\sim& 2.8 \times 10^{49}
\left( \frac{t_{\rm v}}{0.1~\mbox{s}} \right)^{-2}
\left( \frac{E_{\rm ph}}{10^{51}~\mbox{erg}} \right)^2
\left( \frac{\Gamma}{500} \right)^{-7}
\left( \frac{\varepsilon_{\rm p}}{0.5~\mbox{MeV}} \right),
\end{eqnarray}
which is smaller than the number $N_{\rm tot,f}$ adopted
in eq. (\ref{eq:totphe}), but of considerable quantity.
If the condition $N_\pm \ll N_{\rm tot,f}$ is critical for
the model, we need to adjust the model parameters,
e.g., increase of $N_{\rm tot,f}$ by a smaller $B_{\rm f} t_{\rm f}$
or decrease of $N_\pm$ by a larger $\Gamma$.
As shown in Figure \ref{fig:spec}, however,
the photon spectrum may not be the extension of the single power-law
as far as the energy of
$\Gamma \gamma_{\rm th} m_{\rm e} c^2 \sim 26 (\Gamma/500)
(\gamma_{\rm th}/100)$ GeV.
In such cases, the number of the pairs is smaller than the above estimate.
The numerical simulations in \citet{asa09}
is a case where the contribution of the secondary pairs
is negligible, while the pairs injected via the hadronic cascade
dominates in the model of \citet{mur12}.

Another possibility is that the sudden increase of electrons/positrons
due to the non-linear effect
triggers the damping of the turbulence.
As long as the total energy in the turbulence is finite,
an explosive increase of the lepton number causes
a strong dissipation of the turbulence.
In that case, the average energy of the electrons/positrons
in the acceleration process also suddenly decreases.
This leads to inefficiency of gamma-ray emission.
Such a non-linear effect between the turbulence and
secondary pairs may finally determine the duration $t_{\rm f}$.

In conclusion, the turbulence decay time-scale $t_{\rm f}$,
which depends on the electron injection rate including the secondary pairs,
should be much shorter than the dynamical time-scale
in the stochastic acceleration model.
Note that this time-scale is still much longer than the presumed
energy-transfer time-scale from protons to electrons
in the shock acceleration models.

\section{Off-Axis Contribution}
\label{sec:off}

\begin{figure}
	\includegraphics[width=\columnwidth]{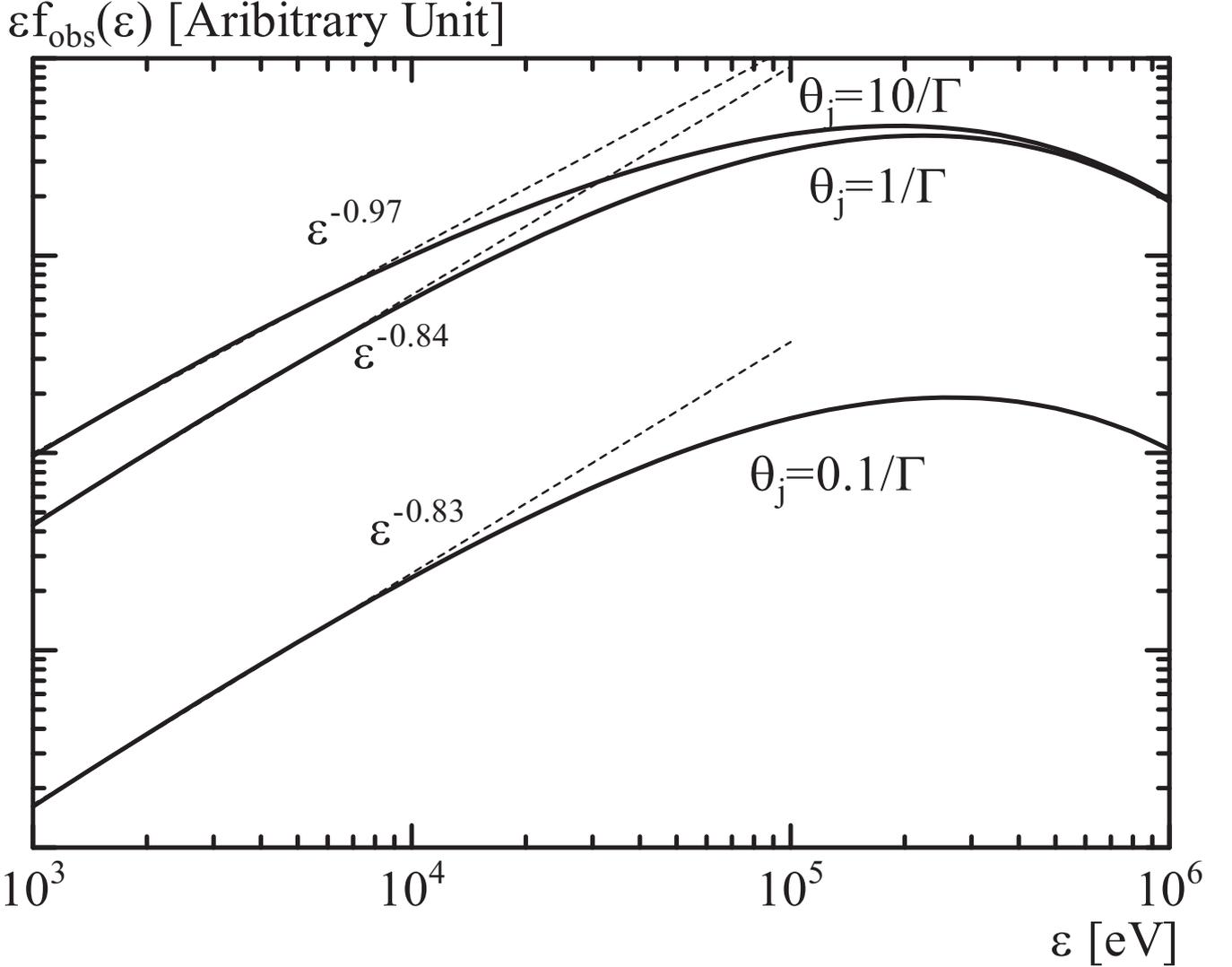}
    \caption{Observable photon spectra for various jet opening angles.}
    \label{fig:off}
\end{figure}

In this section we switch a topic to
focus on the low-energy portion of the Band spectrum.
One may consider that the hard electron spectrum
leads to the ``synchrotron limit'' $\alpha=-2/3$ for the photon index.
Such a hard spectral index is not frequently detected.
However, the best model in Figure \ref{fig:spec} ($b=1$) is well fitted with
the typical Band parameter $\alpha=-1$.
The difference between the model and Band function is hard to
be recognized observationally.
Moreover, the contribution of the off-axis emission can soften the photon spectrum
for observers.
The dominant contribution to photons detected by an observer
is emitted from the jet surface within the opening angle
(relative to the line-of-sight) of $1/\Gamma$
because of the relativistic beaming.
However, the emission from $\theta>1/\Gamma$ can modify the spectrum
as shown in Figure \ref{fig:off}.

Here, we have assumed that photons distributing
isotropically in the rest frame are emitted
from a sphere expanding with $\Gamma=300$.
The photon spectrum in the rest frame has the form
of $F_q$ with $q=5/3$.
We have numerically integrated photons emitted from the surface within
$\theta \leq \theta_{\rm j}$ taking into account
the relativistic beaming effects;
photon energy $\varepsilon'=\varepsilon \Gamma (1-\beta_{\rm sh} \cos{\theta})$,
solid angle $d \Omega=\Gamma^2 (1-\beta_{\rm sh} \cos{\theta})^2 d \Omega'$,
photon number $dN=dN'$, surface $dS=dS'$,
angle $\cos{\theta'}=(\cos{\theta}-\beta_{\rm sh})/
(1-\beta_{\rm sh} \cos{\theta})$,
and $dN'/(d \Omega' dS') \propto |\cos{\theta'}|$
\citep[see][]{asa11},
where $\beta_{\rm sh} \equiv \sqrt{1-1/\Gamma^2}$.

The well-known photon index $-2/3$ for the synchrotron function
\citep{ryb79} is obtained from $F_q(\chi)$ in the limit of $\chi \ll 1$.
Even for the range of $10^{-2} \varepsilon_{\rm p}$--$10^{-1.5} \varepsilon_{\rm p}$,
however, the functions are fitted with a photon index
between $-0.84$ and $-0.79$.
This is seen in the low-energy slope for narrow jet opening angle
of $\theta_j=0.1/\Gamma$ or $1/\Gamma$ in Figure \ref{fig:off}.
For the large opening angle of $10/\Gamma$,
the off-axis contribution slightly enhances
the low-energy flux, which results in a photon index of $\sim -1$.
This slight modification obtained numerically may be hard to
be treated analytically.
If the jet has angle-dependent spectrum like the model
of \citet{lun13}, various photon index would be obtained.

\section{Summary}
\label{sec:sum}

We have discussed the stochastic acceleration by
turbulence in GRB jets.
Since the energy source is the turbulence,
mildly relativistic turbulence is required to
achieve a high radiative efficiency.
When the electron spectral distribution is regulated
by the balance between stochastic acceleration and synchrotron cooling,
the temporal evolutions of the two combinations of the phenomenological parameters,
$N_{\rm tot}/E_0^2$ and $B E_0^2$, determine the final photon spectrum.
If we choose an ideal parameter set,
the photon spectrum can be fitted with the typical Band parameters
$\alpha \sim -1$ and $\beta \sim -2.5$.
The off-axis contribution can soften the low-energy photon index further.
The hard-to-soft evolution of photon spectrum due to the decay of the turbulence
in our model may agree with the observed signatures \citep{nor86,bha94,for95,bur14}.
The duration of the stochastic acceleration should be much shorter
than the dynamical time-scale for fiducial values of the magnetic field.
The injection of the secondary electron--positron pairs may
control the duration time-scale.

The required parameter set is translated into the temporal indices
for the evolutions of the diffusion coefficient,
magnetic field, and total electron number.
The excitation mechanism and evolution of the turbulence in GRB jets
are highly uncertain.
The verifications of the turbulence evolution and details
of the acceleration mechanism are beyond the scope of this paper.
However, the implication obtained from the recent numerical simulations
of decaying MHD turbulence \citep{ino11,zra14,bra15}
seems encouraging for the required parameter evolutions.
We can expect that future simulations of electron acceleration in turbulence,
including the cases with magnetic reconnection \citep{hos12,kag13,dah14,guo14,guo15},
reveal the excitation mechanism and initial condition of the turbulence
to reproduce the GRB emissions.

\section*{Acknowledgments}

We appreciate the anonymous referee for the helpful advice.
This work is partially supported by the Grant-in-Aid for
Scientific Research, No. 25400227 from the MEXT of Japan.







\bsp	
\label{lastpage}
\end{document}